\documentclass[preprint,showpacs,preprintnumbers,nofootinbib]{revtex4}
\usepackage{graphicx}
\usepackage{epstopdf}
\usepackage{dcolumn}
\usepackage{bm}

\newcommand{\nn}{\nonumber}

\newcommand{\be}{\begin{eqnarray}}
\newcommand{\ee}{\end{eqnarray}}
\catcode`\@=11
\def\lsim{\mathrel{\mathpalette\@versim<}}
\def\gsim{\mathrel{\mathpalette\@versim>}}
\def\@versim#1#2{\vcenter{\offinterlineskip
\ialign{$\m@th#1\hfil##\hfil$\crcr#2\crcr\sim\crcr } }}
\catcode`\@=12

\def\thefootnote{\fnsymbol{footnote}}

\begin{document}

\def\thefootnote{\fnsymbol{footnote}}
\begin{flushright}
KANAZAWA-11-12\\
December, 2010
\end{flushright}
\vspace*{0.5cm}

\begin{center}

{\Large \bf 

Triangle Relation of Dark Matter, EDM 
 
and CP Violation in $B^0$ Mixing 

in a Supersymmetric $Q_6$ Model
}

\end{center}

\vspace{0.2cm}

\begin{center}
{\large Yoshiyuki Kaburaki \footnote[1]{e-mail:~kab@hep.s.kanazawa-u.ac.jp},  
Kazuhiro Konya \footnote[2]{e-mail:~konya@hep.s.kanazawa-u.ac.jp},  
Jisuke Kubo\footnote[3]{e-mail:~jik@hep.s.kanazawa-u.ac.jp}}
\vspace {0.5cm}\\
{\it Institute for Theoretical Physics, Kanazawa University,\\
        Kanazawa 920-1192, Japan}
\vspace {0.2cm}\\
and 
\vspace {0.2cm}\\
{\large Alexander Lenz}
\footnote[4]{e-mail:~alexander.lenz@physik.uni-regensburg.de}
\vspace {0.2cm}\\
{\it Institut f{\"u}r Theoretische Physik, Universit{\"a}t Regensburg,\\
        D-93040 Regensburg, Germany}\\
\end{center}
\vspace{0.5cm}

{\Large\bf Abstract}\\
We consider a recently proposed supersymmetric model based on the 
discrete $Q_6$ family group.
Because of the family symmetry and spontaneous CP violation
the electric dipole moment (EDM),
the CP violation in  the mixing of the neural mesons and 
the dark matter mass $m_{\rm DM}$ are closely related. This
triangle relation is  controlled by the size of the $\mu$ parameters.
Loop effects can give rise to 
large contributions to the soft mass insertions, and we find that 
the model allows a large CP violation in the $B^0$ system.
Its size is comparable with the recent 
experimental observations at D0 and CDF, and it could  be observed at LHCb in the
first years. If  the parameter space  is constrained by the neutron EDM, and 
flavor changing neutral currents (FCNC) and CP violations
in  $K^0$ as well as  $B^0$ mixing, 
the triangle relation yields the 
following bound on the 
dark matter candidate: $0.12$ TeV $< m_{\rm DM}< 0.33$ TeV,
which is directly observable at  LHC.
We also compute $a_{sl}^s-a_{sl}^d$, which is observable at 
LHCb, where  $a_{sl}^{s(d)}$
is the semi-leptonic CP asymmetry for the $B_{s(d)} $ system.

\newpage
\setcounter{footnote}{0}
\def\thefootnote{\arabic{footnote}}

\section{Introduction}
Family symmetry is a useful tool \cite{Dine:1993np}-\cite{Babu:2009nn} to suppress 
FCNCs in  supersymmetric (SUSY) extensions of
the standard model (SM) \footnote{For a recent review on family symmetry, 
see \cite{Ishimori:2010au} for instance .}. If it is combined with spontaneous violation
of CP in SUSY models,   CP violation in these models can be suppressed, 
too \cite{Babu:2004tn,Kajiyama:2005rk,Babu:2009nn}.
However, this theoretical idea may be conflict
with the recent measurement of the CP violating dimuon asymmetry $A_{sl}^b$
by the D0 collaboration \cite{Abazov:2010hv}.
Its measured value
$A_{sl}^b = -(9.57 \pm 2.51 \pm 1.46) \cdot 10^{-3} $
is a factor of 42 larger than the SM prediction
$A_{sl}^b = -(2.3^{+0.5}_{-0.6}) \cdot 10^{-4}$ \cite{LN2006},
which has stimulated a number of papers 
\cite{NPdimuon,Kubo:2010mh} dealing with a large CP violation in $B^0$ 
mixing \footnote{For earlier works see e.g. \cite{NPphisold}.}.
Moreover, the CKM fitter group \cite{CKMfitter,Lenz:2010gu}  also obtained from a
global fit to flavor observables a large value for the dimuon asymmetry; 
$A_{sl}^b = -(4.2^{+1.9}_{-1.8}) \cdot 10^{-3}$ \cite{Lenz:2010gu}
\footnote{This value is for the New Physics scenario I
of \cite{Lenz:2010gu}. The UTfit group \cite{UTfit} and Lunghi and Soni 
\cite{Lunghi:2010gv} also reported  large CP violating effects in $B^0$ mixing.}.
If the size of CP violation in a symmetry-based mechanism to suppress CP violation
turns out be of the same order of the SM value above, 
we may be running into a  dilemma between suppressed and large CP violation.
In any case, the mechanism has to take care of small CP violation in  $K^0$ mixing and at 
the same time  allow  large CP violation in  $B^0$ mixing. 
See \cite{Lenz:2010gu} for a large list of references in which diverse theoretical  possibilities 
for large CP violation in $B^0$ mixing have been proposed.

Recently, two of us \cite{Kubo:2010mh} considered a supersymmetric extension of the SM
based on the discrete $Q_6$ family symmetry 
\cite{Babu:2004tn,Kajiyama:2005rk,Kifune:2007fj,Babu:2009nn} 
\footnote{$Q_6$ was considered in past in \cite{Frampton:1995zz}.}.
Due to the family symmetry this model contains  three pairs 
of $SU(2)_L$ doublet Higgs supermultiplets.
We found that  the one-loop effects of the extra Higgs multiplets
to the soft mass insertions
can generically give rise to large contributions to the soft 
mass insertions and that the model allows 
 values for $A_{sl}^b$, that touch the 1 $\sigma$-range of
the fit result from  \cite{Lenz:2010gu}.  
In this paper we will continue with our investigation of this model.
In this model the size of the $\mu$ parameters play
an important role: It enters directly into the above mentioned
one-loop corrections to the soft mass insertions and
into  the EDMs \cite{Babu:2009nn}. If the neutralino LSP should be a dark matter candidate,
then its mass also depends on the $\mu$ parameters.
We are thus particularly interested in the triangle relationship 
between CP violation in $B^0$ mixing,
the EDM and the mass of dark matter candidate.

\section{The model}
We start by considering the superpotential
\be
W
&=&
Y_{ij}^{uI} Q_{i} U_{j}^c  H^u_I+
Y_{ij}^{dI} Q_{i} D_{j}^c  H^d_I+
\mu^{IJ} H^u_I H^d_J~ ,
\label{superP}
\ee
where we have restricted ourselves to
the quark sector and the Higgs sector.
Here  $Q , H^u$ and  $ H^d$ stand for $SU(2)_L$ doublets
of the quark and Higgs supermultiplets,
respectively. The indices $I$ and $J$ indicate different kinds of the Higgs $SU(2)_L$ doublets.
Similarly, $U^c$ and  $D^c$ stand for $SU(2)_L$ singlets 
of the quark supermultiplets.
The structure of the Yukawa matrices
$Y$ and $\mu$ terms are fixed by the $Q_6$ family symmetry\footnote{More details of the model
can be found in \cite{Kajiyama:2005rk,Babu:2009nn}.}.
The $Q_6$ assignment is shown in Table \ref{assignment}, and
\begin{table}
\begin{center}
\begin{tabular}{|c|c|c|c|c|c|c|c|c|c|c|c|}
\hline
 & $Q$ 
 & $Q_3$  
& $U^c,D^c$  
& $U^c_3,D^c_3$ 
 & $L$ & $L_3$ 
 &$E^c,N^c$ & $E_3^c$  &   $N_3^c$ 
  & $H^u,H^d$
 & $H^u_3,H^d_3$ 
\\ \hline
$Q_6$ &${\bf 2}_1$ & ${\bf 1}_{+,2}$ &
 ${\bf 2}_{2}$ &${\bf 1}_{-,1}$ &${\bf 2}_{2}$ &
${\bf 1}_{+,0}$  & ${\bf 2}_{2}$ & 
${\bf 1}_{+,0}$ & ${\bf 1}_{-,3}$ &
${\bf 2}_{2}$ & ${\bf 1}_{-,1}$  \\ \hline
\end{tabular}
\caption{ \footnotesize{The $Q_{6}$ assignment 
of the chiral matter supermultiplets, where  the group theory notation is given in 
 Ref.~\cite{Babu:2004tn}. For completeness we include
 leptons, $L, E^c$ and $N^c$. $R$ parity is also imposed.
}}
\label{assignment}
\end{center}
\end{table}
the  $Q_6$ invariance yields \cite{Babu:2004tn}:
\be
{\bf Y}^{u1(d1)} &=&\left(\begin{array}{ccc}
0 & 0 & 0 \\
0 & 0 & Y_b^{u(d)} \\
0&  Y_{b'}^{u(d)}  & 0 \\
\end{array}\right),~
{\bf Y}^{u2(d2)} =\left(\begin{array}{ccc}
0 & 0 & Y_b^{u(d)}\\
0 & 0 & 0 \\
  -Y_{b'}^{u(d)} &0 & 0 \\
\end{array}\right),\nn\\
{\bf Y}^{u3(d3)}&=&\left(\begin{array}{ccc}
0 & Y_c^{u(d)} & 0\\
Y_c^{u(d)} & 0 & 0 \\
0 &  0 & Y_a^{u(d)} \\
\end{array}\right).
\label{Yuq}
\ee
The only $Q_6$ invariant $\mu$ term is
$(H_1^u H_1^d+H_2^u H_2^d)$, and  no $H_3^u H_3^d$ and 
no mixing between the $Q_6$ doublet and singlet Higgs multiplets  are allowed.
Therefore, there is an accidental global $SU(2)$, implying
the existence of Nambu-Goldstone modes. 
In \cite{BK2010} the Higgs sector is extended to include 
a certain set of SM singlet Higgs multiplets to avoid 
this problem. 
With this extended Higgs sector
one can break the flavor symmetry $Q_6$ and CP invariance
spontaneously. 
Moreover, 
the scalar potential  of the original theory has turned out to have an accidental $Z_2$ invariance
\be
h^{u,d}_+&=&\frac{1}{\sqrt{2}}
(h^{u,d}_1+h^{u,d}_2)
\rightarrow h^{u,d}_+~,~
h^{u,d}_- =\frac{1}{\sqrt{2}}
(h^{u,d}_1-h^{u,d}_2) \rightarrow -h^{u,d}_-,
\label{Hpm}
\ee
where $h$'s are  scalar components of $H$'s.
After  the singlet sector has been integrated out, we obtain
an effective $\mu$ term 
\begin{eqnarray}
W^{\rm eff} &=&
\mu^{++}~(H^{u}_+ H^{d}_+ +H^{u}_- H^{d}_-)+
\mu^{+3}~H^{u}_+ H^{d}_3
+\mu^{3+}~H^{u}_3 H^{d}_+ 
\label{Weff}
\ee
with $H^{u,d}_\pm =
(H^{u,d}_1\pm H^{u,d}_2)/\sqrt{2}$, and the soft-supersymmetry-breaking
Lagrangian
\be
{\cal L}_{\rm soft}^{\rm eff} &=&
m_{H^u}^2~(|h^{u}_+ |^2+|h^{u}_- |^2)+
m_{H^u_3}^2~|h^{u}_3|^2+
m_{H^d}^2~(|h^{d}_+ |^2+|h^{d}_- |^2)+
m_{H^d_3}^2~|h^{d}_3|^2\nonumber\\
& +&\left[~
B^{++}~(h^{u}_+ h^{d}_+ +h^{u}_- h^{d}_-)+
B^{+3}~h^{u}_+ h^{d}_3+
B^{3+}~h^{u}_3h^{d}_+ 
+h.c.\right]~.
\label{Leff}
\end{eqnarray}
( The $A$ terms are suppressed.)
The parameters $\mu$'s and $B$'s  are complex,
they originate from the complex VEVs of the SM singlet Higgs fields of 
the original  theory \cite{BK2010}.
But because of the CP invariance
of the original theory the Yukawa matrices and soft scalar masses are real.
So, the effective superpotential (\ref{Weff})  and the effective 
soft-supersymmetry-breaking Lagrangian (\ref{Leff}) break $Q_6$ and CP softly.
However, thanks to (\ref{Hpm}), 
the VEVs of the form
\be
<h_{-}^{u,d0}> &=& 0~,~ <h_{+}^{u,d0}>
 =\frac{v_{+}^{u,d}}{\sqrt{2}}\exp i \theta_+^{u,d}~, ~
<h_{3}^{u,d0}> 
=\frac{v_3^{u,d}}{\sqrt{2}}\exp i \theta_3^{u,d}
\label{vev1}
\ee
can be realized, where
the $SU(2)$ components of the Higgs fields are defined as
\be
h_I^u &=&(  h_I^{u+}~,~h_I^{u0})~,~
h_I^d=(  h_I^{d0}~,~h_I^{d-})~.
\label{su2-comp}
\ee

To proceed  with our discussion we make a phase rotation of the Higgs superfields
so that their VEVs become real:
$\tilde{H}^{u,d}_{\pm}  = H^{u,d}_{\pm} e^{-i \theta^{u,d}_{+}},~
\tilde{H}^{u,d}_{3} = H^{u,d}_{3} e^{-i \theta^{u,d}_{3}}$.
Then we define
\begin{equation}
\left( \begin{array}{c}
\Phi^{u,d}_L \\\Phi^{u,d}_H \\ \Phi_-^{u,d}
\end{array}
\right)
:= 
\left( \begin{array}{ccc}
 \cos \gamma^{u,d} & \sin \gamma^{u,d} & 0
\\
-\sin \gamma^{u,d} & \cos \gamma^{u,d} & 0
\\
0              & 0             & 1
\end{array}
\right)
\cdot
\left( \begin{array}{c}
\tilde{H}_3^{u,d} \\ \tilde{H}_+^{u,d} \\ \tilde{H}^{u,d}_{-}
\end{array}
\right)~,
\end{equation}
where
\begin{eqnarray}
\cos\gamma^{u,d} &=& v_3^{u,d}/v_{u,d}~,~
\sin\gamma^{u,d} = v_+^{u,d}/v_{u,d}~,~
v_{u,d}=\sqrt{(v_3^{u})^2+(v_+^{u})^2})~.
\label{cosgamma}
\end{eqnarray}
We further define the components of the $SU(2)$ doublet Higgs multiplets as
\begin{eqnarray}
\Phi^u_I &=&\left( \begin{array}{c}
\Phi^{u+}_I\\
\Phi^{u0}_I
\end{array}\right),~
\Phi^d_I =\left( \begin{array}{c}
\Phi^{d0}_I\\
\Phi^{d-}_I
\end{array}\right),~I=L,H,-.
\label{doublet}
\end{eqnarray}
 The light and heavy MSSM-like Higgs scalars  are then given by
\begin{eqnarray}
( v + h-i X )/\sqrt{2} &=&( \phi^{d0}_L)^*\cos\beta
+ (\phi^{u0}_L)\sin\beta~,\label{hX}\nn\\
(H+i A)/\sqrt{2} &=&- (\phi^{d0}_L)^*\sin\beta
+ (\phi^{u0}_L)\cos\beta~,\label{HA}\label{HP}\\
G^+ = - (\phi^{d-}_L)^*\cos\beta &+ & ( \phi^{u+}_L)\sin\beta~,~
H^+ =
 (\phi^{d-}_L)^*\sin\beta+\phi^{u+}_L\cos\beta~,\nn
 \label{CH}
\end{eqnarray}
where 
$X$ and $G^+$ are the Nambu-Goldstone fields, 
$\phi$'s are  scalar components of $\Phi$'s of (\ref{doublet}), and
$v =\sqrt{v_u^2+v_d^2}
~(\simeq 246$ GeV) and $\tan\beta=v_u/v_d$.

\section{The Yukawa sector in the quark mass eigenstates}
The Yukawa sector in the quark mass eigenstates
is needed to compute EDMs mediated by the Yukawa couplings.
The quark mass matrices ${\bf m}^u$ and 
${\bf m}^d$ can be read off from the superpotential
(\ref{superP}) along with (\ref{Yuq}) and (\ref{vev1}). 
Then using the phase matrices defined below
\be
R_L &=&  \frac{1}{\sqrt{2}}\left( \begin{array}{ccc}
1 & 1 & 0\\-1 & 1& 0\\
 0 & 0 &\sqrt{2}
\end{array}\right)~,~
R_R =  \frac{1}{\sqrt{2}}\left( \begin{array}{ccc}
-1 & -1 & 0\\-1 & 1& 0\\
 0 & 0 &\sqrt{2}
\end{array}\right)~,
\label{PR}\\
P_L^q &=&   \mbox{diag.} \left( 
1 ~,~ \exp i2\Delta \theta^{q} ~,~\exp i\Delta \theta^{q}\right)~,
\nn\\
P_R^u &=&(-1)\exp  i \theta_3^u
~\mbox{diag.}\left( 
\exp i2\Delta \theta^{u}~,~ 1~,~\exp i\Delta \theta^{u}\right) ~,
\nn\\
P_R^d &=&\exp  i \theta_3^d
~\mbox{diag.}\left( 
\exp i2\Delta \theta^{d}~,~ 1~,~\exp i\Delta \theta^{d}\right)~,  
\label{dtheta}\\
\Delta \theta^{q} & = &
 \theta_3^q-\theta^q_+~,~(q=u,d)\nn
\ee
 we can bring ${\bf m}^q$ into a real form
${\bf \hat{m}}^q = P_L^{q \dag}R^T_L {\bf m}^q R_R P_R^q$.
The mass matrix ${\bf \hat{m}}^u$ can then be diagonalized as
$O^{uT}_L {\bf \hat{m}}^u O_R^u =
 \mbox{diag.}\left( m_u ~,~ m_c ~,~ m_t\right)$,
and similarly for ${\bf m}^d$, where
$ O_{L,R}^{u,d}$ are orthogonal matrices. 
So, the mass eigenstates $u_{iL}'=(u_{L}', c_{L}',t_{L}')$ 
etc. can be  obtained from $q_L =U_L^q q_L'~,q_R = U_R^q q_R'$,
where
$U_{L(R)}^q =  R_{L(R)} P_{L(R)}^q O_{L(R)}^q$.
Therefore, the CKM matrix $V_{\rm CKM}$ is
 given by 
 \be
 V_{\rm CKM} &=& O^{uT}_L P_L^{u\dag} P_L^d O_L^d=
 O^{uT}_L P_q O_L^d~,
 \label{vckm}
 \ee
 where
  \be
P_q  &=&\mbox{diag.}~(
1, \exp (i2\theta_q), \exp (i\theta_q))~,~
\theta_q =\theta^u_+-\theta^d_+-\theta^u_3
+\theta^d_3~. \label{Pq}
\ee
There are nine independent theory parameters,
which
describe the CKM parameters and the quark masses:
$Y_a^{u,d} v_3^{u,d},
Y_c^{u,d} v_3^{u,d},Y_b^{u,d} v_+^{u,d},
Y_{b'}^{u,d} v_+^{u,d}$ and $\theta_q$.
The set of the theory parameters is thus over-constrained.  Therefore,
there is not much freedom in the parameter space, 
and so it is sufficient to
consider a single point in the space of the theory parameters
of this sector:
\be
Y_a^{u} v_3^{u} &=&1.409 ~m_t ~,
Y_c^{u} v_3^{u}=2.135 \times10^{-4}  ~m_t ~,
Y_b^{u} v_+^{u}=0.0847 ~ m_t ~,
Y_{b'}^{u} v_+^{u}=0.0879 ~ m_t ~,\nn\\
Y_a^{d} v_3^{d} &=&1.258 ~m_b ~,
Y_c^{d} v_3^{d}=-6.037 \times
10^{-3} ~m_b~,Y_b^{d} v_+^{d}=0.0495  ~m_b ~,
Y_{b'}^{d} v_+^{d}=0.6447  ~m_b ~,\nn\\
\theta_q &=& -0.7125 ~.
\label{input}
\ee
With these parameter values we obtain \cite{Araki:2008rn}
\be
m_u/m_t &=& 0.609\times 10^{-5}~,~
m_c/m_t=3.73 \times 10^{-3}~,~
m_d/m_b=0.958 \times 10^{-3}~,\\
\label{ratio}
m_s/m_b & =& 1.69 \times 10^{-2}~,~
| V_{\rm CKM} |= 
\left( \begin{array}{ccc}
0.9740& 0.2266  & 0.00361
\\  0.2264   & 0.9731& 0.0414
  \\ 0.00858 &0.0407& 0.9991
\end{array}\right)~,\\
|V_{td}/V_{ts}| & = &0.211~,~
\sin 2\beta (\phi_1) = 0.695~,~\bar{\rho}=0.152~,~\bar{\eta}=0.343~.
\label{ckm-parameters}
\ee
The mass ratio (\ref{ratio}) is defined at $M_Z$ and consistent with the recent up-dates
of \cite{Xing:2007fb}, and  the CKM parameters above  agree  with those of
Particle Data Group \cite{Amsler:2008zzb} and CKM fitter groups \cite{CKMfitter,UTfit}.
(See \cite{Kubo:2003iw} for the predictions of the model in the lepton sector.)

In the basis of the fermion 
mass eigenstates the Higgs couplings have the following form:
\be
{\cal L}_Y
&=& -\sum_{I=L,H,-}Y_{ij}^{u0I}(~\phi_I^{u0}~)^*
~\overline{u}_{iL}' u_{jR}'+
\sum_{I=L,H,-}Y_{ij}^{d-I}(~\phi_I^{d-}~)^*
~\overline{u}_{iL}' d_{jR}'
\nn\\
& &-\sum_{I=L,H,-}Y_{ij}^{d0I}(~\phi_I^{d0}~)^*
~\overline{d}_{iL}' d_{jR}'
+\sum_{I=L,H,-}Y_{ij}^{u+I}(~\phi_I^{u+}~)^*
~\overline{d}_{iL}' u_{jR}'+h.c.~,
\label{LY}
\ee
where the Higgs fields are defined in (\ref{doublet}),
the Yukawa matrices ${\bf Y}^{u1}$ etc. are given in (\ref{Yuq}), and
\be
{ \bf Y}^{d0L} &=&
 O^{dT}_L R_L^T~ {\bf Y}^{dL}~R_RO^{d}_R
=\sqrt{2} \mbox{diag.} (m_d,m_s,m_b)/v\cos\beta~, \nn\\
{ \bf Y}^{d0H} &=&
 O^{dT}_L R_L^T~ {\bf Y}^{dH}~
 R_R O^{d}_R~,~
  { \bf Y}^{d0-} =\frac{1}{\sqrt{2}}
  O^{dT}_L   R_L^T
   \left({\bf Y}^{d1}-{\bf Y}^{d2}\right)R_R O^{d}_R~,\nn\\
  { \bf Y}^{d-L} &=&
  O^{uT}_L P_q  R_L^T~ {\bf Y}^{dL}~R_R O^{d}_R~,~
{ \bf Y}^{d-H}=
  O^{uT}_L P_q  R_L^T~ {\bf Y}^{dH}~R_R O^{d}_R ~,
  \label{Y-H}\\
    { \bf Y}^{d- -}& =&\frac{1}{\sqrt{2}}
  O^{uT}_L   P_q R_L^T
   \left({\bf Y}^{d1}-{\bf Y}^{d2}\right)R_R O^{d}_R~,\nn\\
{\bf Y}^{dL}&=& \left[\frac{1}{\sqrt{2}}\sin\gamma^d
( {\bf Y}^{d1}+{\bf Y}^{d2})
+\cos\gamma^d  {\bf Y}^{d3}\right] ~,\nn\\
{\bf Y}^{dH}&=& \left[\frac{1}{\sqrt{2}}\cos\gamma^d
( {\bf Y}^{d1}+{\bf Y}^{d2})
-\sin\gamma^d  {\bf Y}^{d3}\right] ~,
\label{YdH}
\ee
and similarly for ${ \bf Y}^{u}$'s, where 
the  matrices other than the Yukawa matrices
are defined  in (\ref{PR}),(\ref{vckm}) and (\ref{Pq}).
One finds that ${\bf Y}^{d0L}$ and  ${\bf Y}^{d0H}$ are real and that 
the only phase  appearing in  ${\bf Y}^{d-L}$  and ${\bf Y}^{d-H}$
is $\theta_q$ given in (\ref{Pq}), which is the same phase entering into
$V_{\rm CKM}$.

\section{Soft mass insertions}

The $A$ terms and
soft scalar mass terms  obey the $Q_6$ family symmetry in the effective theory.
Therefore, the
soft  mass matrices have the  form
\be
{\bf \tilde{m}^2}_{aLL}& =&
{m}^2_{\tilde{a}}~ \mbox{diag.}~
(a_{L}^{a}~,~ a_{L}^{a}~,~ b_{L}^{a})~,~(a=q,l)\nn\\
{\bf \tilde{m}^2}_{aRR}& =&
{m}^2_{\tilde{a}}~ \mbox{diag.}~
(a_{R}^{a}~,~ a_{R}^{a}~,~ b_{R}^{a})~,~(a=u,d,e)
\label{scalarmass}\\
\left({\bf \tilde{m}^2}_{aLR}\right)_{ij} 
&=&
A_{ij}^a\left( {\bf m}^a \right)_{ij}
~,~(a=u,d,e)~\nn
\ee
where ${m}_{\tilde{a}}$ denote the average of the  squark 
and slepton masses, respectively,   $(a_{L(R)}^a, b_{L(R)}^a)$ are
dimensionless free real parameters, 
 $A_{ij}^{a}$ are real free parameters of dimension one,
 and ${\bf m}^a$ are  the respective fermion mass matrices.
According to \cite{Hall:1985dx,Gabbiani:1988rb}
we define the tree-level  supersymmetry-breaking soft mass insertions as
\be
\delta_{LL(RR)}^{a0} &=&
U_{L(R)}^{a\dagger} ~{\bf \tilde{m}^2}_{aLL(RR)}~
 U_{L(R)}^a/{m}^2_{\tilde{a}}~,\\
 \label{DeltaLL}
\delta_{LR}^{u0} &=&
U_{L}^{u\dagger}~\left(~
{\bf \tilde{m}^2}_{uLR}
 -\mu^{IJ} <h^{d0}_J>{\bf Y}^{uJ}
  \right) U_{R}^u/{m}^2_{\tilde{u}}~, \label{DeltaLRu}\\
\delta_{LR}^{d0} &=&
U_{L}^{d\dagger}~ \left(
{\bf \tilde{m}^2}_{dLR}+\mu^{JI} <h^{u0}_J>{\bf Y}^{dJ}
\right) U_{R}^d/{m}^2_{\tilde{d}}~, 
\label{DeltaLRd}
\ee
in the super CKM basis,
where
$U$'s are  unitary matrices that diagonalize the  
quark mass matrices, and 
$h$'s are the neutral Higgs fields defined in (\ref{su2-comp}).
(We restrict ourselves to the quark sector.) The $\mu$ term 
and $A$ term contributions 
to $\delta_{LR}^{u0(d0)}$ are the first and second terms in
(\ref{DeltaLRu}) and (\ref{DeltaLRd}),
respectively. Note that because of the CP invariance,
the A term contributions are real so that only the
$\mu$ term contributes to EDMs.

For the input parameters given in (\ref{input}) we obtain,
for the down quark sector for instance,
\be
(\delta^{d0}_{12})_{LL}
&=&  (\delta^{d0}_{21})_{LL}^*\simeq 
-2.6 \times 10^{-4}~\Delta a_L^{q}~,
~(\delta^{d0}_{13})_{LL}
= (\delta^{d0}_{31})_{LL}^*
\simeq  -8.7\times 10^{-3}~\Delta a_L^q~,\nn\\
(\delta^{d0}_{23})_{LL}
&=& (\delta^{d0}_{32})_{LL}^*
\simeq  -3.0 \times 10^{-2}~\Delta a_L^{q}~,
\label{DeltaLL-d}\\
(\delta^{d0}_{12})_{RR}
&=& (\delta^{d0}_{21})_{RR}^*
\simeq  5.0 \times 10^{-2} ~\Delta a_R^{d} ~,~
(\delta^{d0}_{13})_{RR}
= (\delta^{d0}_{31})_{RR}^*
\simeq -0.10~\Delta a_R^{d}~,\nn\\
(\delta^{d0}_{23})_{RR}
&=& (\delta^{d0}_{320})_{RR}^*
\simeq  0.39~\Delta a_R^{d}~,\nn
\ee
where
$\Delta a_{L}^{q} = a_{L}^{q}-b_{L}^{q}~,~
\Delta a_{R}^{d} =a_{R}^{d}-b_{R}^{d}.$
The A term contributions to the left-right insertions  are
\be
(\delta^{d0}_{12})_{LR}(A)
&\simeq &
1.9 (\tilde{A}_{1}^{d}-\tilde{A}_{2}^{d})\times 10^{-5} ~,~
(\delta^{d0}_{21})_{LR}(A)
\simeq 
(- 2.2\tilde{A}_{1}^{d}+1.7  \tilde{A}_{2}^{d} )\times 10^{-5}~,\nn\\
(\delta^{d0}_{13})_{LR}(A)
&\simeq &
(1.0 \tilde{A}_{1}^{'d}+4.0 \tilde{A}_{2}^{'d}) \times 10^{-5}~,~
(\delta^{d0}_{31})_{LR}(A)
\simeq 
5.8 \tilde{A}_{2}^{d} \times 10^{-4}~,
\nn\\
(\delta^{d0}_{23})_{LR}(A)
&\simeq &
1.4 \tilde{A}_{2}^{'d}\times 10^{-4}~,~
(\delta^{d0}_{32})_{LR}(A)
\simeq 
-2.3 \tilde{A}_{2}^{d} \times 10^{-2} ~,
\label{DeltaLRA}\\
(\delta^{d0}_{12})_{LR}(A)
&\simeq &
1.9 (\tilde{A}_{1}^{d}-\tilde{A}_{2}^{d})\times 10^{-5} ~,~
(\delta^{d0}_{21})_{LR}(A)
\simeq 
(- 2.2\tilde{A}_{1}^{d}+1.7  \tilde{A}_{2}^{d} )\times 10^{-5}~,\nn\\
(\delta^{d0}_{13})_{LR}(A)
&\simeq &
(1.0 \tilde{A}_{1}^{'d}+4.0 \tilde{A}_{2}^{'d}) \times 10^{-5}~,~
(\delta^{d0}_{31})_{LR}(A)
\simeq 
5.8 \tilde{A}_{2}^{d} \times 10^{-4}~,
\nn\\
(\delta^{d0}_{23})_{LR}(A)
&\simeq &
1.4 \tilde{A}_{2}^{'d}\times 10^{-4}~,~
(\delta^{d0}_{32})_{LR}(A)
\simeq 
-2.3 \tilde{A}_{2}^{d} \times 10^{-2} ~,\nn
\ee
where $\tilde{A}_{i}^{d} ~(\tilde{A}_{i}^{'d})
=[A_{i}^{d}~(\tilde{A}_{i}^{'d})]/m_{\tilde{d}}]
[0.5 ~\mbox{TeV}/m_{\tilde{d}}]$, and the real parameters $A_{i}^{d}$ and
$A_{i}^{'d}$ represent four
independent elements of $A_{ij}^{d}$ given in  (\ref{scalarmass}).
The $\mu$ term contributions can be obtained from 
the second terms of (\ref{DeltaLRu}) and (\ref{DeltaLRd}).

The mass insertions above are the tree-level ones.
In \cite{Kubo:2010mh} it has been shown that the one-loop corrections to
them, especially to $(\delta^{d0}_{ij})_{LL}$, can be large in the presence
of more than one pair of the Higgs $SU(2)_L$ doublet.
Moreover, it has been found that in the present model
the one-loop corrections are needed
to obtain a large CP violation in  $B^0$ mixing
that are comparable with the observations at Tevatron.
These one-loop corrections depend strongly on the
parameters in the Higgs sector, and we use the formula
given in \cite{Kubo:2010mh} to do the numerical analysis in the last section.

\section{Dark matter, EDM and $B^0$ mixing}
\subsection{LSP and Dark matter}
We assume that the LSP is a neutralino and is a dark matter candidate
in this model.
Because of $Z_2$ defined in (\ref{Hpm}) the higginos can also be grouped
into the $Z_2$ even and odd sectors
\footnote{Since $Z_2$ is not an exact symmetry of the theory,
the even and odd states will mix with each other in higher orders in perturbation theory.}.
The higginos in the $Z_2$ odd sector have no mixing with the gauginos.
If therefore the LSP belongs to the  $Z_2$ odd sector, the LSP is
a pure higgsino state with the mass $\mu^{++}$. 
For this LSP to be a dark matter candidate, $\mu^{++}$ has to be 
larger than $O(1)$ TeV and at the same time smaller than
the other $\mu$'s and gaugino masses. This parameter region can
not satisfy  the EDM constraint without an extreme
fine tuning  because we need  relatively small $\mu$'s
to satisfy the EDM constraint in the present model \cite{Babu:2009nn}.
So, we may assume that the LSP belongs to the $Z_2$ even sector.
The mass matrix of the neutralinos in $Z_2$ even sector is
\begin{eqnarray}
{\bf M}_{N\rm even}^F &=&\left( \begin{array}{cccccc}
M_1 & 0 & s_W s_\beta M_Z & -s_W c_\beta M_Z  & 0 & 0 \\
0 & M_2 & -c_W s_\beta M_Z & c_W c_\beta M_Z  & 0 & 0 \\
 s_W s_\beta M_Z & -c_W s_\beta M_Z  & 0& -\mu_L & 0 &-\mu_{LH}\\
 -s_W c_\beta M_Z & c_W c_\beta M_Z  &   -\mu_L & 0 &
 -\mu_{HL} & 0\\
0 & 0 &  0 &  -\mu_{HL} & 0&-\mu_{H} \\
0 & 0 &  -\mu_{LH}   & 0 &  -\mu_{H} & 0  \\
\end{array}\right)~,
\label{MFeven}
\end{eqnarray}
where $c_\beta=\cos\beta, ~c_W=\cos\theta_W$, and similarly for 
$s_\beta$ and $s_W$ ($\theta_W$ is the Weinberg angle).
Because of the EDM constraint we expect the mass of the LSP is
relatively light $O(\mbox{few} 100)$ GeV. Therefore, 
the LSP has to be  a mixture of the higginos and the gauginos
to obtain a desirable relic density $\Omega h^2 \simeq 0.11$.
So, we require that the gaugino fraction of the LSP is 
in a range between $65$\% and $95$\% (see for instance 
\cite{Jungman:1995df}), and assume that
if this is satisfied, the neutralino LSP can be a dark matter candidate in the present model.

\subsection{EDM}
Our concern here is the neutron EDM, $d_n$, because the electron EDM
in this model is extremely suppressed \cite{Babu:2009nn}.
There are two sources for $d_n$: the Yukawa sector because of
the multi Higgs structure  and 
the SUSY breaking sector \footnote{See for instance \cite{Pospelov:2005pr}.}.
Here we simply assume that $d_n$ can be obtained
from
$d_n = \frac{1}{3}(4d_d-d_u)$, where  $d_{u(d)}$ is the EDM of the u(d) quark.
The experimental upper bound is given 
by \cite{Amsler:2008zzb}
\be
d_n/e &\lsim& 6.3 \cdot 10^{-26}~\mbox{cm}~.
\label{constedm}
\ee

\subsubsection{Yukawa contribution}
We start in  the Yukawa sector.
The one-loop diagrams can be
divided into: the photon is attached to a quark or a charged Higgs, and
the internal Higgs is neutral or charged \cite{Deshpande:1993py}.
The contribution  to $d_n/e$ with the neutral  Higgs boson exchange 
(satisfying the constraint (\ref{constMH}) )
 is less than $O(10^{-31})$ cm as was previously found in \cite{Babu:2009nn}.
We have computed the contribution with the charged  Higgs boson 
exchange and found that it is slightly smaller  than the upper bound
(\ref{constedm}). The result indeed depends on the mass of the heavy Higgs
bosons. However, as we will argue in the last section, the heavy Higgs masses can not
be  freely increased in this model. Therefore, this model predicts 
$d_n/e$ which is close to the upper bound (\ref{constedm}).

\subsubsection{SUSY breaking contribution}
The second source is the SUSY breaking sector.
To obtain $d_n$ we use the approximate result of
\cite{Gabbiani:1996hi} which takes into account only the 
gluino contribution
\be
d_d/e &=& -\frac{2 \alpha_s}{9\pi}\xi~
\mbox{Im} (\delta^{d0}_{11})_{LR}~,~
d_u/e = \frac{4 \alpha_s}{9\pi}\xi~
\mbox{Im} (\delta^{u0}_{11})_{LR}~,
\label{doe}
\ee
where we have assumed that $m_{\tilde{g}}=m_{\tilde{u}}
=m_{\tilde{d}}=m_{\tilde{q}}$, and 
$\xi \simeq 0.12$ is the QCD correction \cite{Polchinski:1983zd,Deshpande:1993py}.
Since the $A$'s are real, only the $\mu$ terms contribute  to
$\mbox{Im} (\delta_{11}^{u(d)0})_{LR}$, and therefore,
\be
\mbox{Im} (\delta^{u0}_{11})_{LR}&=&\frac{1}{\tan\beta}~\mbox{Im} (\mu_{L}) m_u+
 \frac{v\cos\beta}{\sqrt{2}}~\mbox{Im} (\mu_{HL})~{\bf Y}^{u0H}_{11}
/{m}^2_{\tilde{u}}~,\label{ImDeltaLRu}\\
\mbox{Im} (\delta^{d0}_{11})_{LR}
&=&  \tan\beta~\mbox{Im} (\mu_{L}) m_d+
\frac{v\sin\beta}{\sqrt{2}}~\mbox{Im} (\mu_{LH})
~{\bf Y}^{d0H}_{11}
/{m}^2_{\tilde{d}}~,
\label{ImDeltaLRd}
\ee
where we have used (\ref{DeltaLRu}) and  (\ref{DeltaLRd}), and 
${\bf Y}^{u0H}$ and ${\bf Y}^{d0H}$ are defined in (\ref{Y-H}).
In the last section the equations (\ref{ImDeltaLRu}) and  (\ref{ImDeltaLRd}) 
will be used to relate the dark matter mass $m_{\rm DM}$, the neutron EDM and
the CP violation in  $B^0$ mixing.

\subsection{$B^0$ mixing}
The tree-level contributions to the $B^0$ mixing coming from the heavy
neutral Higgs boson exchange in this model
are small if
\be
\cos\beta M_H  &\gsim& 1.2~\mbox{ TeV}~,
\label{constMH}
\ee
is satisfied \cite{Kifune:2007fj,Babu:2009nn},
where $M_H^2$ is the  $(\varphi_H^d-\varphi_H^d)$ element 
of the inverse of the mass squared matrix of the neutral Higgs bosons
in $Z_2$ even sector ($\varphi_H^d$ is the scalar component of 
$\Phi_H^{d0}$ given in (\ref{doublet})).
In the following discussion we assume this, so that the only relevant 
contribution comes from the SUSY breaking sector.
Therefore, the total matrix element  $M_{12}^q$ 
in the neutral meson mixing can be written as
$M_{12}^q =M^{\rm SM,q}_{12} + M^{\rm SUSY,q}_{12}$,
where $M^{\rm SM,q}_{12}$ and 
 $M^{\rm SUSY,q}_{12}$ are the SM contribution and the  SUSY contribution,
 respectively.  We take into account only
 the  dominant contribution (gluino exchange) for
 $ M^{\rm SUSY,q}_{12}$ given in
\cite{Gabbiani:1996hi}.
(See e.g. \cite{Gorbahn:2009pp} for a more refined calculation)

We follow \cite{LN2006} to parameterize new physics effects as
\be
M^{\rm  SM,q}_{12} + M^{ \rm SUSY,q}_{12}
&=&  M^{\rm SM,q}_{12} \cdot \Delta_q\, ,
\label{M12}
\ee
and 
consider 
$\Delta M_{q}$ and the flavor specific CP-asymmetry 
$a_{sl}^{q}$ in terms of the complex number 
$\Delta_{q} = |\Delta_{q}| e^{i \phi^{\Delta}_{q}}$, where $q=d,s$, and
\be
\Delta M_q  & = & 2| M^{\rm SM,q}_{12}| \cdot |\Delta_q | ~,~
\nn\\
a_{sl}^q
& =&
 \frac{| \Gamma^q_{12} |}{| M^{\rm SM,q}_{12} |}
\cdot \frac{\sin \phi_q}{|\Delta_q|}~,
\phi_q=\phi_q^{\rm SM} + \phi^\Delta_q~.
\label{afs} 
\ee
The SM values are given e.g.  in \cite{LN2006}, in which the results of
\cite{BBD06,BBGLN98,BBGLN02,BBLN03,rome03} are used:
\be
2~M^{{ \rm SM},d}_{12}&=&
0.56(1\pm 0.45)\exp (i 0.77)
~\mbox{ps}^{-1}~,
\label{phis}\\
 2~M^{{ \rm SM},s}_{12} &=&
20.1(1\pm 0.40)\exp (-i 0.035)
~\mbox{ps}^{-1}~,\nn\\
\phi_d^{\rm SM}& = &(-0.091 ~{}^{+0.026}_{-0.038})~\mbox{rad}~,~
\phi_s^{\rm SM}=(4.2\pm 1.4) \cdot10^{-3}~\mbox{rad}~,\nn
\ee
where the errors are dominated by the uncertainty in the 
decay constants and bag parameters
\footnote{Note that the values for $M^{{ \rm SM},q}_{12}$ quoted above  are those 
in the standard parameterization of the CKM matrix 
\cite{Amsler:2008zzb} and that the CKM matrix
obtained from (\ref{vckm}) 
is not in the standard parameterization. Therefore, we have to 
express the supersymmetric contribution $M^{{\rm SUSY},q}_{12}$
in the standard parameterization of the CKM matrix 
before actual calculations.}.
We use the central values of (\ref{phis}) for our calculations,
while requiring the (conservative) constraints 
\be
0.6 < & \frac{\Delta M_{d,s}}{\Delta M_{d,s}^{\rm exp }}  < 1.4 ~,~
\frac{2 | M^{\rm SUSY,K}_{12} |}{\Delta M_{K}^{\rm exp }} <2~,~
\frac{\mbox{Im} M^{\rm SUSY,K}_{12}\lambda_u^2}{\sqrt{2} 
\Delta M_K^{\rm exp }|\lambda_u|^2}
 <  \epsilon_K=2.2\cdot 10^{-3}~,
\label{const2}
\ee
where
$\lambda_u= (V_{\rm CKM})^*_{us} (V_{\rm CKM})_{ud}$.

The same sign dimuon asymmetry $A_{sl}^b$ measured at
D0  \cite{Abazov:2010hv} is 
a linear combination of the semileptonic CP-asymmetries in the $B_d$ and 
in the $B_s$ system:
\begin{equation}
A_{sl}^b = (0.494\pm0.043) \cdot a_{sl}^s + (0.506\pm0.043) \cdot a_{sl}^d \, .
\label{Asl}
\end{equation}
The SM value for $A_{sl}^b$ is given by
$A_{sl}^b = -(2.3^{+0.5}_{-0.6}) \cdot 10^{-4}$
 \cite{LN2006}, while the fit result yields \cite{Lenz:2010gu}
\be
A_{sl}^b &=& -(4.2^{+1.9}_{-1.8}) \cdot 10^{-3}~.
\label{asl-fit}
\ee

\section{Result and conclusion}
Most of the free parameters belong to the Higgs sector
and the SUSY breaking sector. The parameter space is so large
that it will be beyond the scope of the
present paper to analyze the complete parameter space.
Instead, we first look for a benchmark point in the parameter space
that satisfies  all the requirements (\ref{constedm}), (\ref{constMH}),
(\ref{const2}) and (\ref{asl-fit}).
Then we consider  neighbor points and look for a border
beyond which the constraints are no longer simultaneously satisfied.
The border is extended by a certain amount and 
the parameter space to be considered is defined as 
such that  is surrounded by the extended border.

Note that a larger  $\tan\beta$ means a smaller
 $\cos\beta$ which requires a finer fine tuning in the
 Higgs sector  in order to satisfy (\ref{constMH}). 
$\tan\beta=10$ for instance would require $M_H \gsim 12$ TeV.
In the following analysis we consider   a benchmark value $\cos\beta=0.3~
(\tan\beta \simeq 3.18)$, which implies $M_H \gsim 4$ TeV.
Further, $\Delta a_{L,R}^{q,d}$ in (\ref{DeltaLL-d}) are $O(1)$ free parameters.
We assume that $|\Delta a_{L,R}^{q,d}| \lsim 15$.

We start with the dark matter mass $m_{\rm DM}$ (the mass of the neutralino LSP).
It is the smallest eigenvalue of (\ref{MFeven}) and depends
on the gaugino masses and $\mu$ parameters.
The $\mu$ parameters directly enter into EDM
 (see (\ref{ImDeltaLRu}) and (\ref{ImDeltaLRd}) ), while
the tree-level mass insertions $(\delta_{ij}^{d0})_{RR,LL}$ given in 
(\ref{DeltaLL-d}) do not depend on the $\mu$ parameters.  However, their
one-loop corrections do depend on them \cite{Kubo:2010mh}.
So, the dark matter mass $m_{\rm DM}$ in the present model is constrained 
by  EDM and  by the  mixing of the neutral meson systems.
We find that $m_{\rm DM}$ is indeed bounded above and below:
\be
0.12 ~\mbox[TeV] \lsim m_{\rm DM}  ~\lsim 0.33 ~\mbox[TeV] ~,
\label{mDM}
\ee
 where  we have required 
(\ref{const2}) and  (\ref{asl-fit}) with $\cos\beta M_H\simeq 1.2$  TeV
and used $m_{\tilde{g}}=m_{\tilde{u}}
=m_{\tilde{d}}=m_{\tilde{q}}=0.5$ TeV.
The upper bound becomes larger if the size of the $\mu$ parameters increases.
However, the size of the second term in the rhs  of   (\ref{DeltaLRd}), in particular
for $(\delta_{32}^{d0})_{LR}$, increases, too. The upper bound given in (\ref{mDM}) 
corresponds to $|(\delta_{32}^{d0})_{LR}|\sim O(10^{-2})$ which is about the upper limit
to satisfy the constraint from $b \to s \gamma$ \cite{Gabbiani:1996hi}
\footnote{$|(\delta_{23}^{d0})_{LR}|$ is two orders of magnitude smaller than 
$|(\delta_{32}^{d0})_{LR}|$.}. 
Similarly, if we increase $\cos\beta M_H$, the one-loop effect
becomes larger because of a larger SUSY breaking in the extra Higgs sector,
and consequently (\ref{const2}) will be violated.
To reduce the one-loop effect, we have  to increase the size of the $\mu$ parameters
to reduce the SUSY breaking.
But this was not allowed because of the $b \to s \gamma$ constraint.
Therefore, (\ref{mDM})
should  be regarded as the area  of $m_{\rm DM}$ of the present model.
The phenomenological feature of the dark matter of the present model is basically the
same as the one of the MSSM. Therefore, it could be observed in various  future
experiments \cite{Altunkaynak:2008ry}.

\begin{figure}[htbp]
\begin{tabular}{cc}
\begin{minipage}{0.5\hsize}
\begin{center}
\includegraphics*[width=1\textwidth]{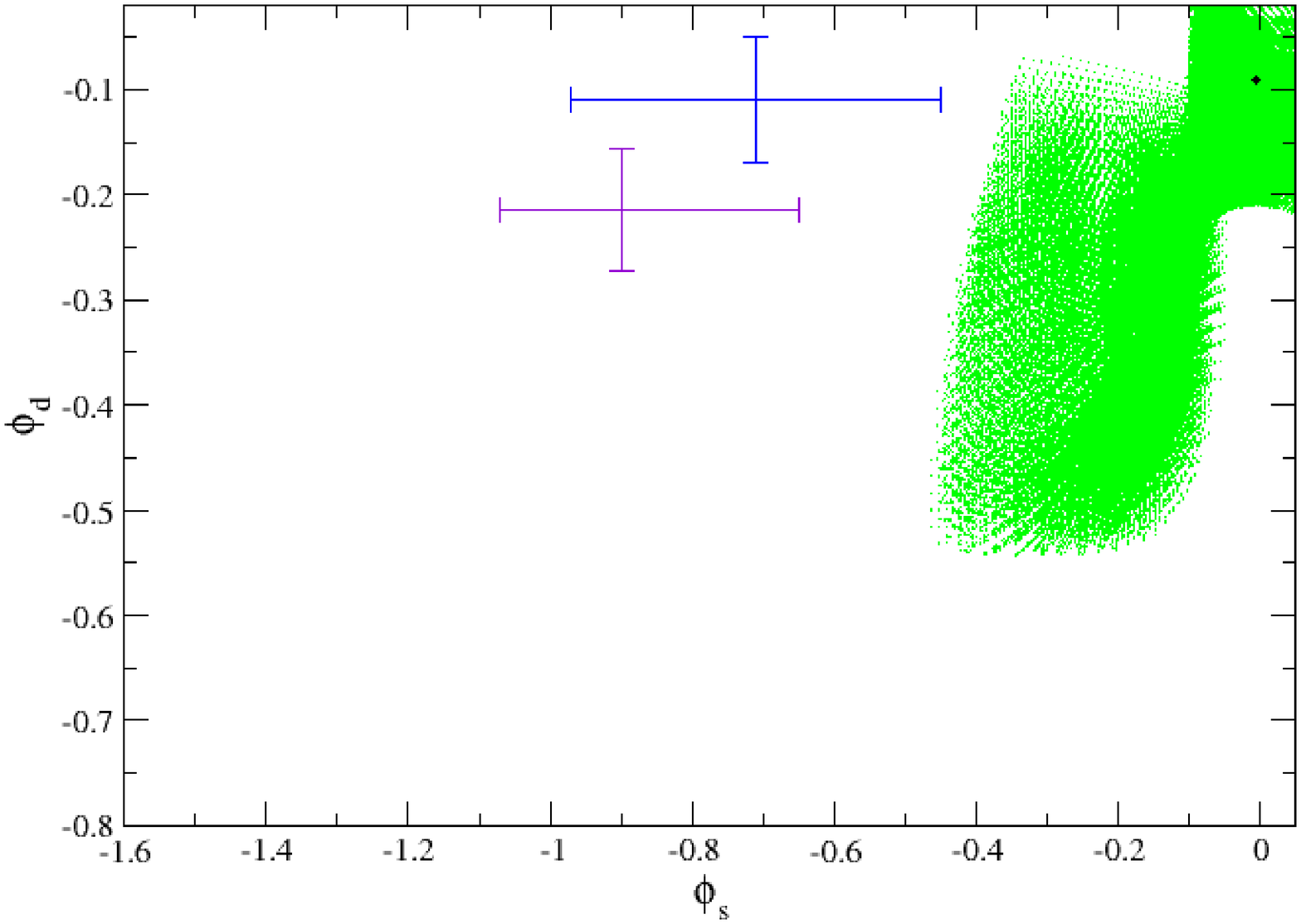}
\caption{\label{dphis-dphid}{\scriptsize
The prediction in the $\phi_s - 
\phi_d$ plane.
The fit result of the CKMfitter  group   (purple)  \cite{Lenz:2010gu}
and that  of the UTfit group  (blue)  \cite{UTfit} are also shown.
The black dot is the SM value.}}
\end{center}
\end{minipage}
\begin{minipage}{0.5\hsize}
\begin{center}
\includegraphics*[width=1\textwidth]{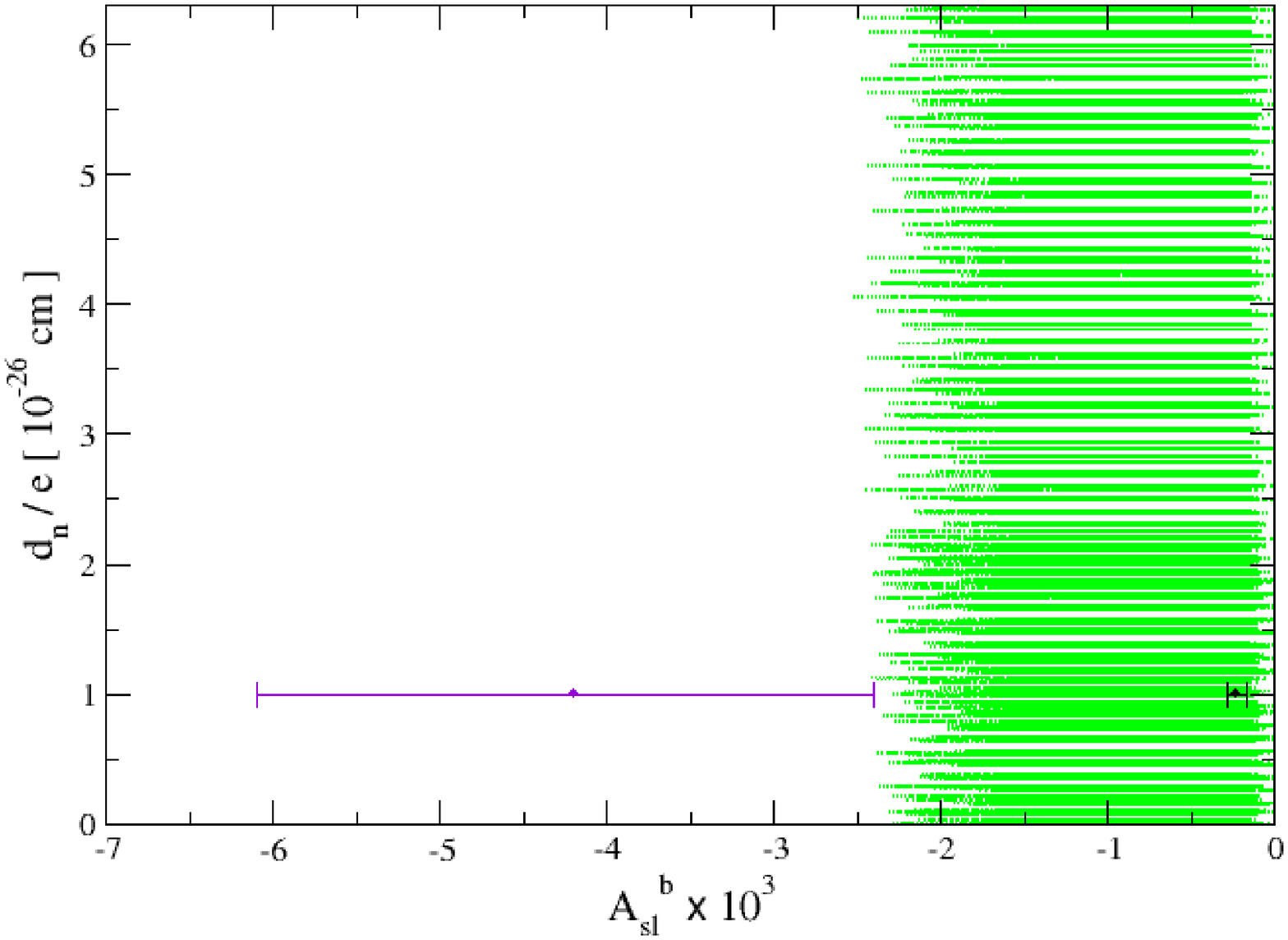}
\caption{\label{Ab-edm}{\scriptsize
The same sign dimuon asymmetry $A_{sl}^b$ against
$d_n/e$.
The fit result for $A_{sl}^b$ 
is $ -(4.2^{+1.9}_{-1.8}) \cdot 10^{-3}$  (purple) \cite{Lenz:2010gu},
and the D0  result \cite{Abazov:2010hv} is 
$A_{sl}^b = -(9.57 \pm 2.51 \pm 1.46) \cdot 10^{-3} $.
The SM value is shown in  black.
}}
\end{center}
\end{minipage}
\end{tabular}
\end{figure}

Next we consider the extra phases $\phi_s$ and $\phi_d$ 
defined in  (\ref{afs}), which are shown in Fig.~\ref{dphis-dphid}.
Also shown are the fit results of the CKMfitter  group   (purple)  \cite{Lenz:2010gu}
and  the UTfit group  (blue)  \cite{UTfit}.
As we see from the figure,
the theoretical values  are comparable with 
the fit values and about one order of magnitude larger than the
SM value (black dot).
The same sign dimuon asymmetry $A_{sl}^b$ against
$d_n/e$ is shown in Fig.~\ref{Ab-edm}.
A large imaginary part of the $\mu$ parameters, on one hand, 
produces a large CP violation in  $B^0$ mixing. On the other hand, 
the large imaginary part implies a large EDM.
Fig. ~\ref{Ab-edm} shows that 
the SUSY contribution to $d_n$  in this model   can be
made very small,
while allowing a large $A_{sl}^b$ which in magnitude is comparable
with the fit result (\ref{asl-fit}).
As we see from Fig. ~\ref{Ab-edm} the error
in $A_{sl}^b$ is very crucial to test the prediction of the model.
We hope that the error will be reduced by the future experiments.

\begin{figure}[htb]
\includegraphics*[width=0.5\textwidth]{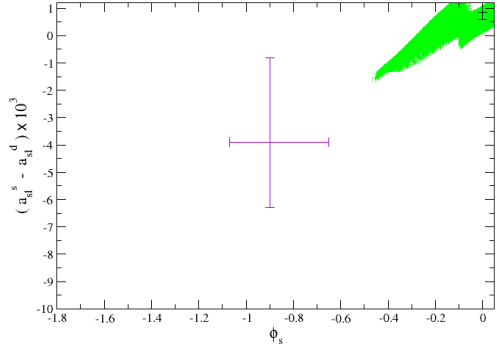}
\caption{\label{asl-asd}\footnotesize
The prediction of $a_{sl}^s-a_{sl}^d$, where the horizontal axis stands for 
$\phi_s$.
The fit result   for $a_{sl}^s-a_{sl}^d$
is $ -(3.9^{+2.4}_{-3.1}) \cdot 10^{-3}$  (purple) , while
the SM value  is $ (0.793^{+0.066}_{-0.214}) \cdot 10^{-3}$ (black).
}
\end{figure}
In Fig.~\ref{asl-asd} we plot the prediction of $a_{sl}^s-a_{sl}^d$
against $\phi_s$. This combination of the asymmetries can be measured
at LHCb, and the experimental sensitivity with one f$\mbox{b}^{-1}$,
which will be achieved in 2011\cite{Teubert:2010dc}, is sufficient
to test it.

\vspace*{5mm}
J.~K. is partially supported by a Grant-in-Aid for Scientific
Research (C) from Japan Society for Promotion of Science (No.22540271).

\end{document}